\documentclass[runningheads]{llncs}

\usepackage{latexsym,amssymb,amsmath}
\usepackage{xspace}
\usepackage{comment}
\usepackage{url}
\usepackage{enumerate}
\usepackage[defblank]{paralist}
\usepackage{graphicx}
\graphicspath{{imgs/}}
\usepackage{hyperref}
\usepackage{subfig}
\usepackage{microtype}
\usepackage{color}
\usepackage{todonotes}
\usepackage{times}

\newif\ifdraft
\draftfalse

\newcommand{\mf}{\mathfrak}

\newcommand{\myi}{\emph{(i)}\xspace}
\newcommand{\myii}{\emph{(ii)}\xspace}
\newcommand{\myiii}{\emph{(iii)}\xspace}

\renewcommand{\mf}[1]{\Upsilon}

\title{Medical Process Modeling: \\an Artifact-Centric Approach \\ {\normalsize Position Paper}}

\author{Dmitry Solomakhin}
\authorrunning{Dmitry Solomakhin}

\institute{
 KRDB Research Centre,
 Free University of Bozen-Bolzano, \\
 Piazza Domenicani 3, 39100 Bolzano, Italy\\
 \email{solomakhin@inf.unibz.it}
}

\begin{document}

\maketitle
\sloppy

\setcounter{footnote}{0}

\begin{abstract}
In this position paper we argue that just as traditional business process modeling has been adopted to deal with clinical pathways, also the artifact-centric process modeling technique may be successfully used to model various kinds of medical processes: physiological processes, disease behavior and treatment processes. We also discuss how a proposed approach may be used to deal with an interplay of all the processes a patient is subject to and what are the queries that might be imposed over an overall patient model.

%
\end{abstract}

{\small \textbf{Keywords:} healthcare process modeling, knowledge representation, artifact-centric \\ \phantom{Keywordssssss:} systems, formal verification, process compliance.}

\section{Introduction}
\label{intro}

Recent development of information technology has significantly affected the way how a healthcare industry operates.
Medical Information Systems have advanced from Electronic Health Record solutions to sophisticated clinical decision support systems and clinical pathways management systems. While the former are usually rely on evidence-based machine learning and hypothesis generation, the latter is mostly based on workflow technologies, originating from Business Process Modeling (BPM). 


A common drawback of classical healthcare process modeling approaches is being \emph{activity-centric}: they mainly focus on the control-flow perspective, lacking the connection between the treatment process and the patient data manipulated during its execution. Therefore, such approaches are more focused on curing a particular disease, rather than treating a concrete patient. Moreover, whenever a patient suffers from several diseases at the time, it becomes particularly difficult to align several clinical pathways, since the combinations of those might lead to life-threatening conflicts difficult to trace over time. 

To tackle the similar problem in knowledge-intensive business domains, the artifact-centric paradigm has recently emerged as an approach in which processes are guided by the evolution of business data objects, called \emph{artifacts} \cite{Nigam03:artifacts,CH09}. A key aspect of artifacts is coupling the representation of data of interest, called \emph{information model}, with \emph{lifecycle constraints}, which specify the acceptable evolutions of the data maintained by the information model. Current research in artifact-centric business process management mirrors the development of traditional business process formalisms. 
On the one hand, new modeling notations are being proposed to tackle artifact-centric processes. A notable example is the Guard-State-Milestone (GSM) graphical notation \cite{Damaggio:2011:EIF:2040283.2040315}, which corresponds to way executive-level stakeholders conceptualize their processes \cite{Bhatt-2007:artifacts-customer-engagements}.
On the other hand, formal foundations of the artifact-centric paradigm are being investigated in order to capture the relationship between processes and data and support formal verification  \cite{Deutsch:2009:AVD:1514894.1514924,BeLP12,DBLP:journals/corr/abs-1203-0024}. 
This form of reasoning support is particularly important in the artifact-centric setting, due to the subtle interactions between the data and process components. 

In this position paper we argue that just as traditional business process modeling has been adopted to deal with clinical pathways, also the artifact-centric process modeling technique may be successfully used to model various kinds of medical processes: physiological processes, disease behavior and treatment processes. We also discuss how a proposed approach may be used to deal with an interplay of all the processes a patient is subject to and what are the queries that might be imposed over an overall patient model.

The rest of the paper is organized as follows. Section 2 provides an overview of artifact-centric systems and the current state-of-art. Section 3 introduces the proposed approach to medical process modeling. Section 4 concludes the paper with a discussion.

\section{Overview of artifact-centric systems}
\label{sec-acs}
The foundational character of artifact-centric systems is the combination of static properties, i.e., the data of interest, and
dynamic properties of a business process, i.e., how it
evolves. \emph{Artifacts}, the key entities of a given domain, are characterized by \myi an
\emph{information model} that captures domain-relevant data, and
\myii a \emph{lifecycle model} that specifies how the artifact
progresses through the process.
In this paper, we focus on the Guard-Stage-Milestone (GSM) approach for
artifact-centric modeling, recently proposed by IBM \cite{Damaggio:2011:EIF:2040283.2040315}.
In this section, we introduce the full expressive power of the GSM methodology, using its original business-related terminology and a relevant example, while in the next section we narrow down the approach to appligycation in the medical processes domain. 
\subsection{Guard-Stage-Milestone modeling}
GSM is a declarative modeling framework that has been designed with the goal of being executable and at the same time enough high-level to result intuitive to executive-level stakeholders. 
The GSM information model uses (possibly nested) attribute/value pairs to capture the
domain of interest.
The key elements of a  lifecycle model are \emph{stages},
\emph{milestones} and \emph{guards}. 
Stages are (hierarchical) clusters of
activities (\emph{tasks}), intended to update and extend the
data of the information model. They are associated to milestones, operational objectives to be achieved when the stage is under execution. Guards control the activation of stages and, like milestones, are described in terms of data-aware expressions, called
\emph{sentries}, involving events and conditions over the artifact
information model. Sentries have the form $[\textbf{on } e \textbf{ if }
cond]$, where $e$ is an event and $cond$ is a condition
over data. Both parts are optional, supporting pure event-based or
condition-based sentries.
Tasks represent the atomic units of work. Basic tasks are used to
update the information model of some artifact instance (e.g., by using
the data payload associated to an incoming event). Other tasks are
used to add/remove a nested tuple. A specific
\emph{create-artifact-instance} task is instead used to create a new
instance of a given artifact type; this is done by means of a two-way
service call, where the result is used to create a new tuple for the
artifact instance, assign a new identifier to it, and fill it with the
result's payload. Obviously, another task exists to remove a given
artifact instance.
In the following, we use \emph{model} for the intensional level of a specific process described in GSM, and \emph{instance} to denote a
GSM model with specific data for its information model.

The execution of a business process may involve several
\emph{instances} of artifact types described by a GSM model. At any
instant, the state of an artifact instance (\emph{snapshot}) is stored
in its information model, and is fully characterised by: \myi values of attributes in the data model, 
\myii status of its stages (open or closed) and
\myiii status of its milestones (achieved or invalidated).
%
Artifact instances may interact with the external world by exchanging
typed \emph{events}. In fact, \emph{tasks} are considered to be
performed by an external agent, and their corresponding execution is
captured with two event types: a \emph{service call},
whose instances are populated by the data from information model and then sent to the environment; 
and a \emph{service call return}, whose instances represent the
corresponding answer from the environment and are used to incorporate
the obtained result back into the artifact information model.
The environment can also send unsolicited (one-way) events, to trigger specific guards or milestones. 
Additionally, any change of a status attribute, such as opening a stage or  achieving a milestone, triggers an internal event, which can be further used to govern the artifact lifecycle.

\begin{example}
\label{ex:gsm}
\small
\begin{figure}[t]
\centering
\includegraphics[width=.9\textwidth]{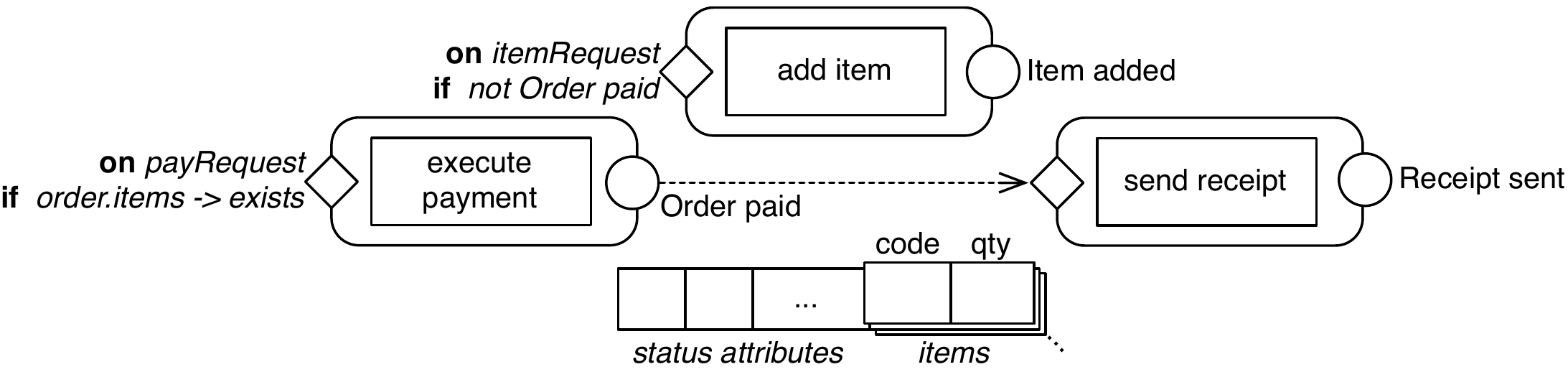}
\caption{GSM model of a simple order management process\label{fig:order-gsm}}
\end{figure}
Figure \ref{fig:order-gsm} shows a simple order management process
modeled in GSM. The process centers around an \emph{order} artifact, whose
information model is
characterized by a set of status attributes (tracking the status of
stages and milestones), and by an extendible set of ordered
\emph{items}, each constituted by a code and a quantity. The order
lifecycle contains three top-level atomic stages (rounded rectangles), respectively used to
manage the manipulation of the order, its payment, and the delivery of
a payment receipt. The order management stage contains a task
(rectangle) to add items to the order. It opens every time an $itemRequest$ event is received, 
provided that the order has not yet been paid. This is represented using a
logical condition associated to a guard (diamond). The stage closes when the task is executed, by achieving an ``item added''
milestone (circle). 
A payment can be
executed once a $payRequest$ event is issued, provided that the order
contains at least one item (verified by the OCL condition
$order.items \rightarrow exists$). As soon as the order is paid, and the
corresponding milestone achieved, the receipt delivery stage is
opened. This direct dependency is represented using a dashed arrow,
which is a shortcut for the condition $\textbf{on } Order~paid$, representing
the internal event of achieving the ``Order paid'' milestone. 
\end{example}

\subsection{Operational semantics of GSM}
GSM is associated to three well-defined, equivalent execution semantics, which
discipline the actual enactment of a GSM model \cite{Damaggio:2011:EIF:2040283.2040315}.
Among these, the \emph{GSM incremental semantics} is based on a form
of Event-Condition-Action (ECA) rules and is centered
around the notion of \emph{GSM Business steps} (\emph{B-steps}). An artifact instance remains idle until it receives an incoming event from the environment. It is assumed that such events arrive in a sequence and get processed by artifact instances one at a time. A B-step then describes what happens to an \emph{artifact snapshot} $\Sigma$, when a single incoming event $e$ is incorporated into it, i.e., how it evolves into a new snapshot $\Sigma'$ (see Figure 5 in \cite{Damaggio:2011:EIF:2040283.2040315}).

%
%

The evolution of a GSM system composed by several artifacts can be
described by defining the initial state (initial snapshot of all
artifact instances) and the sequence of event instances generated by
the environment, each of which triggers a particular B-step, producing
a sequence of system snapshots. This perspective intuitively leads to
the representation of a GSM model as an infinite-state transition
system, depicting all possible sequences of snapshots supported by
the model. The initial configuration of the information model represents
the initial state of this transition system, and the incremental semantics provides the actual
transition relation. The source of infinity relies in the payload of
incoming events, used to populate the information model of artifacts
with fresh values (taken from an infinite/arbitrary domain). Since
such events are not under the control of the GSM model, the system must be
prepared to process such events in every possible order, and with
every acceptable configuration for the values carried in the payload.
The analogy to transition systems opens the possibility of using a
formal language, e.g., a (first-order variant of) temporal logic, to verify whether the GSM
system satisfies certain desired properties and requirements. For
example, one could test generic correctness properties, such as
checking whether each milestone can be achieved (and each stage will
be opened) in at least one of the
possible systems' execution, or that whenever a stage is opened, it
will be always possible to eventually achieve one of its
milestones. Furthermore, the modeler could also be interested in
verifying domain-specific properties, such as checking whether for the
GSM model in Figure~\ref{fig:order-gsm} it is possible to obtain a receipt before the payment is processed.

\section{Artifact-centric medical processes}
\label{sec-medproc}
In this section we show how the artifact-centric process modeling technique may be used in order to model various kinds of medical processes a patient is involved in. As a matter of fact, it is the patient, or his Electronic Health Record (EHR), what becomes the key entity of the domain, i.e. the main \emph{artifact type}. The current state of the patient is described by a set of attributes, each of which corresponds to a particular entry in the EHR. Such attributes may either have a certain atomic datatype (e.g. blood pressure, oxygen level, etc) or be of a semantic nature (e.g. \emph{having a headache}, \emph{being pregnant}, etc). In the latter case, such semantic attributes may be borrowed either from an existing or from a specifically developed ontology. We then model all the relevant medical processes as a lifecycle of the \emph{Patient} artifact type. In particular, we distinguish the following types of processes \footnote{Although one could also model aforementioned processes on the molecular level, we assume that they are described on the semantic level. As a matter of fact, models of both granularities may coexist, since processes on molecular level may be modeled as subprocesses on the semantic level.}: 

\begin{itemize}
\item \emph{Physiological processes}. These are the processes that every human is subject to: digestion, breathing (oxygen metabolism), blood circulation, other physiological dependencies.
\item \emph{Disease behavior}. Each disease may, in fact, be described by a set of processes, which model different phases of the disease and its influences on various physiological processes. 
\item \emph{Treatment processes}. By these processes we understand both different treatment procedures (e.g. surgery) and drug treatment. While the former may be described by the patient-data-driven workflows, the latter is usually described by pharmacokinetics and pharmacodynamics of the particular drug. 
\end{itemize}

We then model each of the aforementioned processes in the artifact-centric manner, where such processes are described as the evolution of the patient data along the execution of such process. However, the specific character of the processes at hand implies several peculiarities: 
\begin{itemize}
\item Since we model patient data as the main artifact type, we can safely exclude the \emph{create-artifact-instance} service calls since that will imply ``creation'' of new patient.
\item For the similar reason we consider only incoming one-way events triggered by the environment, which may correspond to updating test results or exposure to some external reagents (e.g. poisons).
\item We also consider only basic tasks, performed by an external agent (e.g. surgeon, nurse, etc), which  update the current state of patient data. 
\end{itemize} 

\begin{figure}[t]
\centering
\includegraphics[width=.9\textwidth]{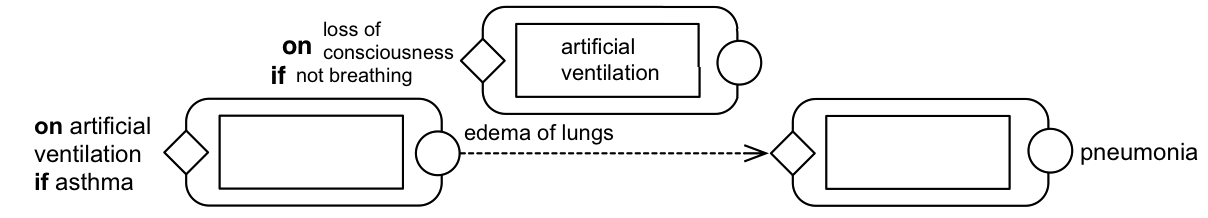}
\caption{GSM model fragment of a simple clinical case\label{fig:health-gsm}}
\end{figure}

Having modeled all physiological, disease and treatment processes one could obtain very interesting inferences and deductions over such model. For instance, analyzing the process depicted on Figure \ref{fig:health-gsm}, one could infer that using an artificial ventilation as treatment procedure for loss of consciousness in case of asthma may potentially lead to a specific kind of pneumonia \footnote{All the examples are purely fabricated and may be not aligned to real clinical cases.}. 
Another potential usage of such process model originates from the operational semantics of GSM. Verification of a reachability property for the resulting transition system may be used to obtain the sequence of specific actions, treatments and other external events, which may be used to achieve the desired state of the patient. For instance, one could formulate the following query over an instance of a model (i.e. specific patient data):
\begin{center}
\emph{Having the current state of this particular patient, are there any drug treatments that will result in healing the diabetes but will not result in arrhythmia?}
\end{center}

\section{Discussion}
\label{sec-discussion-relwork}
In this work we have given an insight on how an emerging methodology for business process modeling may be used to model various medical processes. 
Applying this methodology to healthcare processes allows to benefit from already existing mechanism of the formal verification of the GSM artifact-centric paradigm for a very rich first-order temporal logic, tailored to the artifact-centric setting \cite{DBLP:journals/corr/abs-1203-0024}.
Moreover, such approach eliminates a so-called ``data and process engineering divide'', which affects many contemporary process-aware medical information systems and due to which such systems are more focused on curing a particular disease, rather than treating a concrete patient. 

The results presented in our paper can be used to generalize this approach towards more complex models.
Future work towards application of the artifact-centric modeling to healthcare processes also includes further studies of the domain of interest in order to determine relevant usecases and required queries for further assessment of the approach.

\bibliographystyle{splncs03}

\begin{thebibliography}{1}
\providecommand{\url}[1]{\texttt{#1}}
\providecommand{\urlprefix}{URL }

\bibitem{BeLP12}
Belardinelli, F., Lomuscio, A., Patrizi, F.: An abstraction technique for the
  verification of artifact-centric systems. In: Proc. of KR. AAAI Press (2012)

\bibitem{Bhatt-2007:artifacts-customer-engagements}
Bhattacharya, K., Caswell, N.S., Kumaran, S., Nigam, A., Wu, F.Y.:
  Artifact-centered operational modeling: {Lessons} from customer engagements.
  IBM Systems Journal  46(4),  703--721 (2007)

\bibitem{CH09}
Cohn, D., Hull, R.: Business artifacts: A data-centric approach to modeling
  business operations and processes. IEEE Data Eng. Bull.  32(3) (2009)

\bibitem{Damaggio:2011:EIF:2040283.2040315}
Damaggio, E., Hull, R., Vaculin, R.: On the equivalence of incremental and
  fixpoint semantics for business artifacts with guard-stage-milestone
  lifecycles. Information Systems  (2012)

\bibitem{Deutsch:2009:AVD:1514894.1514924}
Deutsch, A., Hull, R., Patrizi, F., Vianu, V.: Automatic verification of
  data-centric business processes. In: Proc.\ of ICDT. pp. 252--267. ICDT '09,
  ACM (2009)

\bibitem{DBLP:journals/corr/abs-1203-0024}
Hariri, B.B., Calvanese, D., Giacomo, G.D., Deutsch, A., Montali, M.:
  Verification of relational data-centric dynamic systems with external
  services. CoRR  abs/1203.0024 (2012)

\bibitem{Nigam03:artifacts}
Nigam, A., Caswell, N.S.: Business artifacts: An approach to operational
  specification. IBM Systems Journal  42(3) (2003)

\end{thebibliography}
\newcommand{\SortNoOp}[1]{}

\end{document}